\documentclass[epj]{webofc}
\usepackage[varg]{txfonts}   
\woctitle{CNR*13}
\begin{document}
\title{Coupled-channel optical model potential for rare earth nuclei}
%
%

\author{M. Herman\inst{1}\fnsep\thanks{\email{mwherman@bnl.gov}} \and
        G. P. A. Nobre\inst{1}\fnsep\thanks{\email{gnobre@bnl.gov}} \and
        A. Palumbo\inst{1} \and
        F. S. Dietrich\inst{2} \and
        D. Brown\inst{1} \and
        S. Hoblit\inst{1}
}

\institute{National Nuclear Data Center, Brookhaven National Laboratory, Upton, NY 11973-5000, USA
\and
P.O. Box 30423, Walnut Creek, CA, 94598, USA
          }

\abstract{%
The global spherical optical model by Koning and Delaroche is generalized to enable its use in coupled-channel calculations on well deformed nuclei in the rare-earth region.   The generalization consists in adding the coupling of the ground state rotational band, deforming the potential by introducing appropriate quadrupole and hexadecupole deformations and correcting the optical model potential radius to preserve volume integral of the spherical optical potential. We choose isotopes of three rare-earth elements (W, Ho, Gd), which are known to be nearly perfect rotors, to perform a consistent test of our conjecture on integrated cross sections as well as on angular distributions for elastic and inelastic neutron scattering.  The only additional input are experimentally determined deformations, which we employ without any adjustments.  The results are clearly superior compared to the spherical optical model calculations with dramatic improvement at low incident energies. 
}
\maketitle
\section{Introduction}
\label{intro}

The coupled-channel theory is a natural way of accounting for nonelastic channels, in particular those arising
from collective excitations. Proper treatment of such excitations is often essential
to the accurate description of reaction experimental data. Optical potentials (OP) needed for coupled-channels calculations are normally obtained through proper parametrization and parameter fitting in order to reproduce experimental data sets for a specific nucleus. Such phenomenological OP's  might achieve very good  agreement with experimental data, as they were specifically adjusted to do so, but do not lend themselves to extrapolation to other nuclei, unless  there are explicit measurements that allow to readjust OP individually. Therefore, the predictive power of the coupled-channels method is hampered by  the lack of reliable OP for the nuclei with no or scarce experimental data.   On the other hand, there are quite reliable spherical OP's applicable to spherical or slightly-deformed nuclei.  Therefore,   developing a method capable of employing such well-tested global OP's to the particular case of deformed nuclei would be desirable.

Inspired by the recent work by Dietrich et al., substantiating the validity of the adiabatic assumption in coupled-channel calculations, we explore the possibility of generalizing (deforming) a global spherical optical model potential to make it usable in coupled-channel calculations on statically deformed nuclei.

Optical potentials (OP) have been widely used to describe nuclear reaction data by implicitly accounting for the
effects of excitation of internal degrees of freedom and other nonelastic processes. 
An OP is called global when this fitting process is consistently done for a variety of nuclides.

There are global spherical OP's that have been fit to nuclei below and above the region of statically deformed
rare-earth nuclei, but these potentials have been viewed as inappropriate for use in coupled-channels calculations,
since they do not account for the loss of flux through the explicitly included inelastic channels.  On the other
hand, a recent paper \cite{Dietrich:2012} shows that scattering from rare earth and actinide nuclei is very near
the adiabatic (frozen nucleus) limit, which suggests that the loss of flux to rotational excitations might be
unimportant.  In this paper we test this idea by performing coupled channel calculations with a global spherical
optical potential by deforming the nuclear radii but making no further adjustments.  We note an alternative
approach (Kuneida \emph{et al.} \cite{Kunieda:2007}), which has attempted to unify scattering from spherical and
deformed nuclei by considering all nuclei as statically deformed, regardless of their actual deformation.

This work corresponds to a preliminary attempt to extend the approach initially presented in Ref.~\cite{NobreND2013}, focusing on angular distributions for the cases of neutron scattered by Gd, Ho, and W nuclei.

\section{Adiabatic model for rare-earths}

Due to the high moment of inertia and consequent low excitation energies of the ground-state band members of the statically deformed nuclei in the rare-earth region, the deformed nuclear configuration may be regarded as ``frozen'' during the scattering. This means that all the internal degrees of freedom not associated with the strong deformation may assumed to be accounted for by a spherical optical potential that describes well the nuclei in the neighboring region, in an adiabatic approach. Therefore, the only channels that need to be treated explicitly (e.\ g., through couple-channel methods) are the ones arising from the static deformation.

The spherical OP that was deformed in our coupled-channel calculations was the global Koning-Delaroche (KD)~\cite{KD}, unmodified except for a small change in the radius parameters to ensure volume conservation when the nucleus is deformed.  Since the KD potential describes scattering from nuclei both above and below the deformed rare earth region very well, we make the assumption that the imaginary potential adequately describes the internal nuclear excitations in the rare earths also.  This  picture is consistent with the adiabatic approximation.  The coupled channel calculations account for the external (rotational) excitations of the target.  These assumptions are tested in the calculations shown in this paper.

The process of deforming a spherical OP to explicitly consider collective excitations within the couple-channel framework is done in the standard way of replacing the radius parameter $R$ in each Woods-Saxon form factor by the angle dependent expression:
\begin{equation}
\label{Eq:DefRadius}
R(\theta)=R_0\left( 1+\sum_\lambda{\beta_\lambda Y_{\lambda0}(\theta)} \right)
\end{equation}
where $R_0$ is the undeformed radius of the nucleus, and $\beta_\lambda$ and $Y_{\lambda0}(\theta)$ are the deformation parameter and spherical harmonic for the multipole $\lambda$, as seen in Ref.~\cite{Krappe1976}, for example. The deformed form factor obtained from Eq.~\ref{Eq:DefRadius} is then expanded in Legendre polynomials numerically.

\begin{figure}
\centering
  \includegraphics[height=.3\textheight, clip, trim= 5mm 4mm 5mm 14mm]{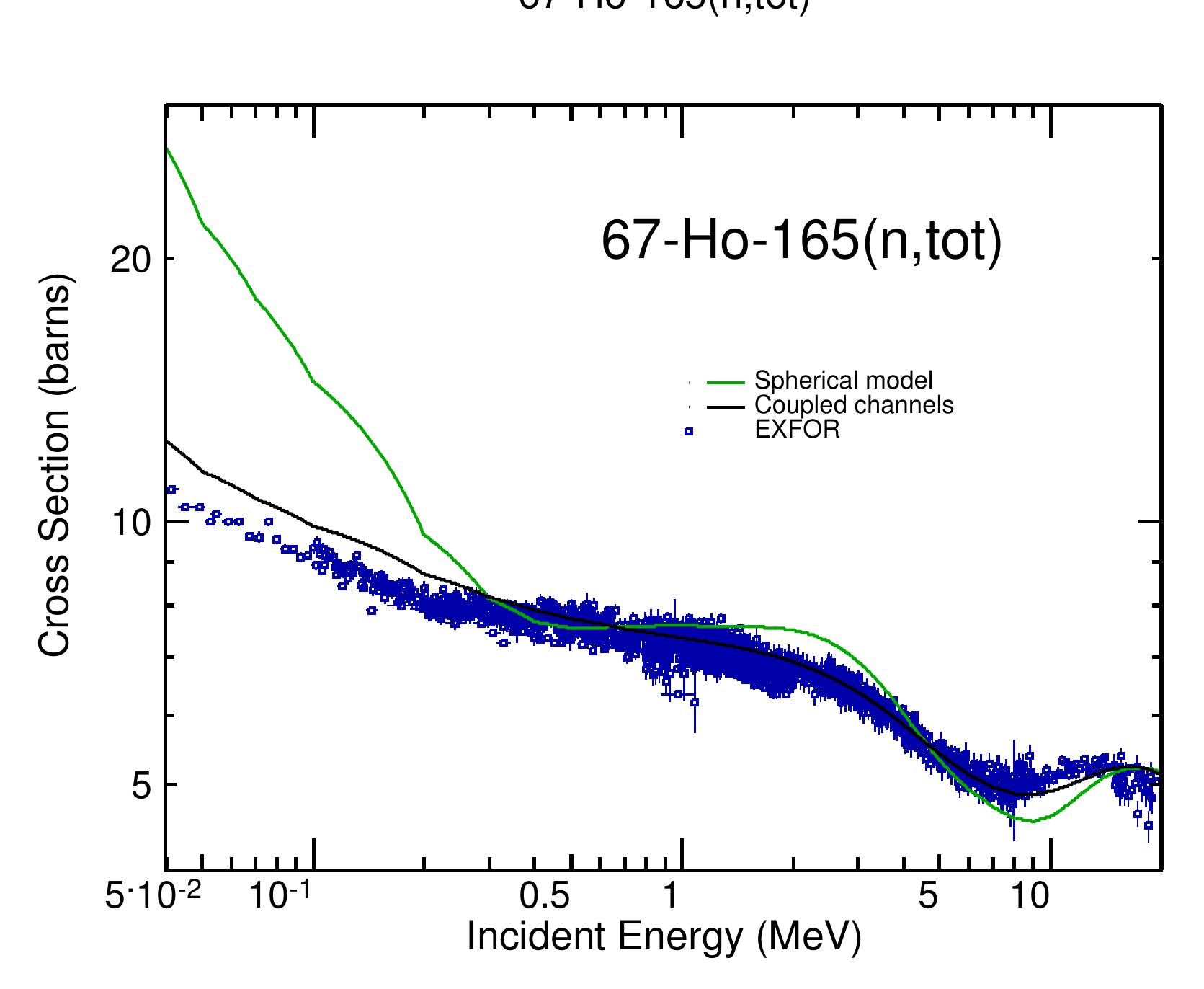}
  \caption{Total cross sections for neutron-induced reaction on $^{165}$Ho. The black curve corresponds to coupled-channel calculations within our model, while the green curve indicates, for comparison purposes, the result from a spherical model calculation. Experimental data taken from EXFOR \cite{EXFOR}.}
  \label{Fig:Total}
\end{figure}

We use in our calculations the \textsc{Empire} reaction code \cite{Herman:2007,EmpireManual}, in which the direct reaction part is calculated by the code \textsc{Ecis} \cite{Raynal70,Raynal72}. In order to test our model we perform coupled-channel calculations, coupling the ground state rotational band, for neutron-incident reactions on selected rare-earth nuclei, namely $^{152,154}$Sm, $^{153}$Eu, $^{155,156,157,158,160}$Gd,
$^{159}$Tb, $^{162,163,164}$Dy, $^{165}$Ho,  $^{166,167,168,170}$Er, $^{169}$Tm, $^{171,172,173,174,176}$Yb, $^{175,176}$Lu,
$^{177,178,179,180}$Hf, $^{181}$Ta, and $^{182,183,184,186}$W. All those nuclides have at least 90 neutrons, indicating static deformation, therefore making them suitable candidates for interpolation through the adiabatic limit.     We then compared, as an initial test, the obtained coupled-channel results for total cross sections with plain spherical calculations with the undeformed KD optical potential. In this initial step, only quadrupole deformations were considered, having their values taken from the compilation of experimental values from Raman \emph{et al.} \cite{Raman}.  The overall result is a dramatic improvement in the agreement  with experimental data, in particular in the lower neutron-incident energies. Fig.~\ref{Fig:Total} clearly illustrates the very good description of observed total cross section in the case of $^{165}$Ho, obtained through our coupled-channel model. 

In carrying out the calculations, it is important to couple a sufficient number of rotational states to achieve convergence, and we have carried out tests to ensure this. Such analysis is shown in Ref.~\cite{Nobre-RTFNB}, where it is also demonstrated that this convergence can be energy-dependent.

\subsection{Compound-nucleus observables}

After the initial success in describing direct-reaction quantities, such as total cross sections, we analyzed the model predictions for observables that depend also on the compound-nucleus decay. The models adopted to describe the emissions from the compound nucleus were basically default options in \textsc{Empire} code, which means standard Hauser-Feshbach model with properly parametrized Enhanced Generalized Superfluid Model (EGSM) level densities \cite{fade}, modified Lorentzian  (version 1)  $\gamma$-ray strength functions \cite{plu01,plu02,plu03}, width fluctuation correction up to 3 MeV in terms of the HRTW approach \cite{HRTW,HHM}, and with transmission coefficients for the inelastic outgoing channels also calculated within coupled-channel approach. Pre-equilibrium was calculated within the exciton model~\cite{Griffin:66}, as based on the solution of the
master equation~\cite{Cline:71} in the form proposed by Cline~\cite{Cline:72}
and Ribansky~\cite{Ribansky:73} (using \textsc{Pcross} code \cite{Herman:2007,EmpireManual}) with mean free path multiplier set to 1.5.

\begin{figure}
  \includegraphics[height=.24\textheight, clip, trim=     4mm 3mm 5mm 3mm]{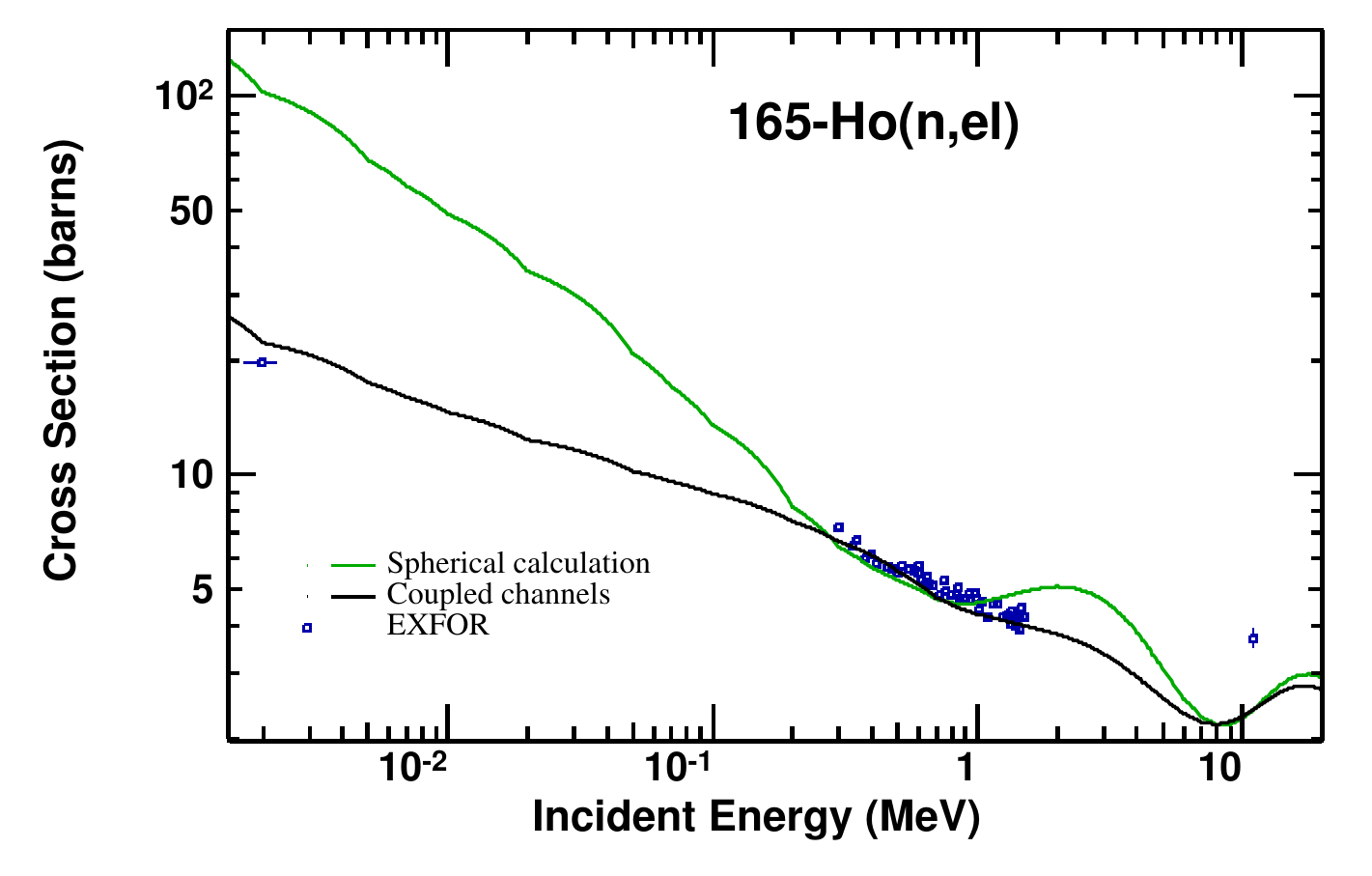} \hspace{-1.8mm}
  \includegraphics[height=.24\textheight, clip, trim=   24mm 3mm 5mm 3mm]{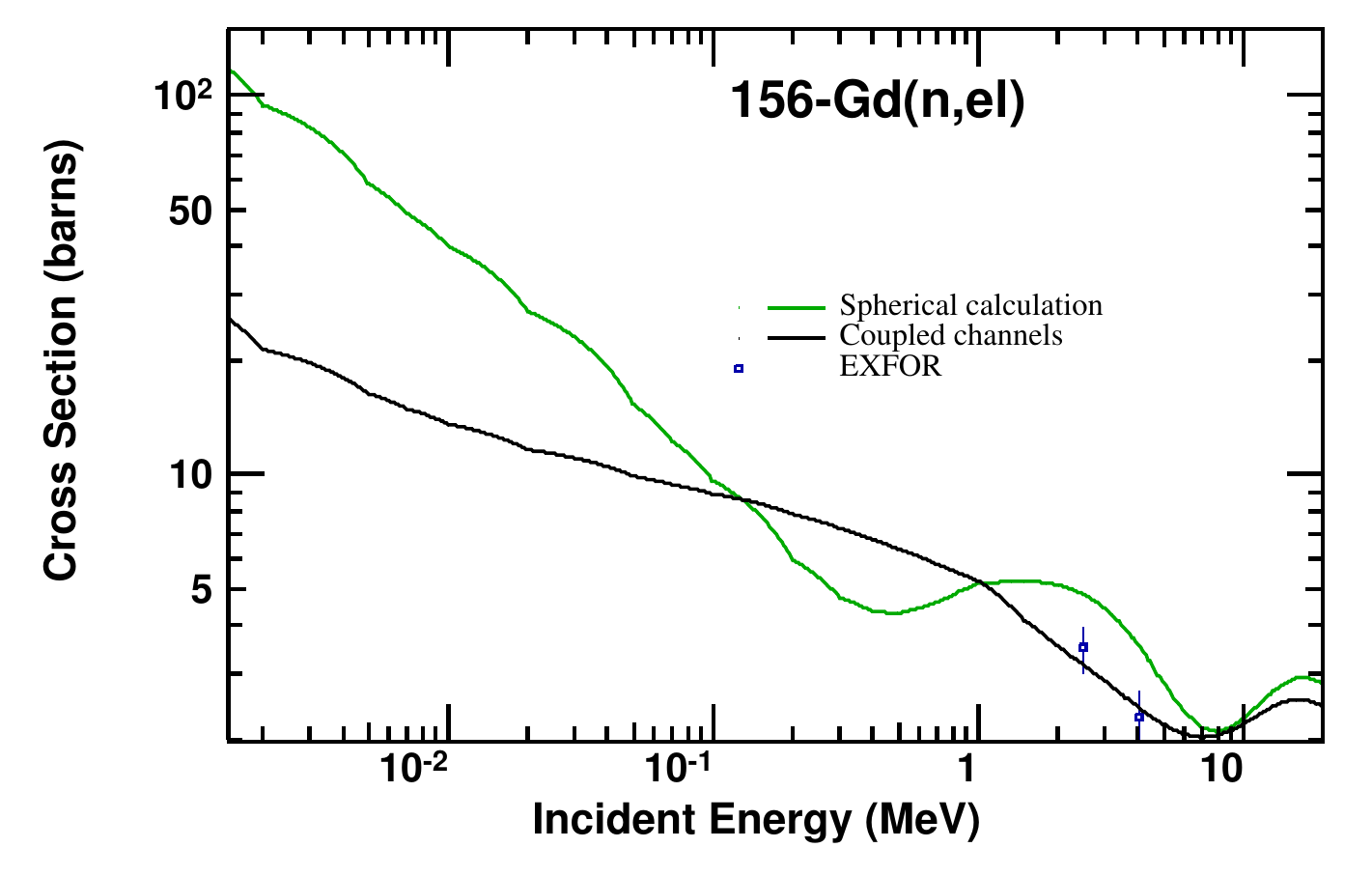}
  \caption{Angle-integrated elastic cross sections for the case of  $^{165}$Ho (left panel) and $^{156}$Gd (right panel) targets. Black curves correspond to predictions by our coupled-channel model while green curves were obtained by spherical model calculations. Experimental data taken from EXFOR \cite{EXFOR}.}
  \label{Fig:Elastic}
\end{figure}

 In Fig.~\ref{Fig:Elastic}, as an example, we compare with experimental data the angle-integrated elastic cross sections for incident neutrons on $^{165}$Ho and $^{156}$Gd obtained by our coupled-channel calculations. Even though there are not as many data available as in the case of total cross sections (Fig.~\ref{Fig:Total}), the low-energy point ($\sim$ 3 keV) for $^{165}$Ho (Fig.~\ref{Fig:Elastic}, left panel) and the few high energy points ($\gtrsim$ 2 MeV) for $^{156}$Gd (Fig.~\ref{Fig:Elastic}, right panel) indicate again a very good agreement between our couple-channel model and the experimental data (in contrast with the spherical-model calculations).

\begin{figure}
  \includegraphics[height=.22\textheight, clip, trim=     27mm 28mm 32mm 119mm]{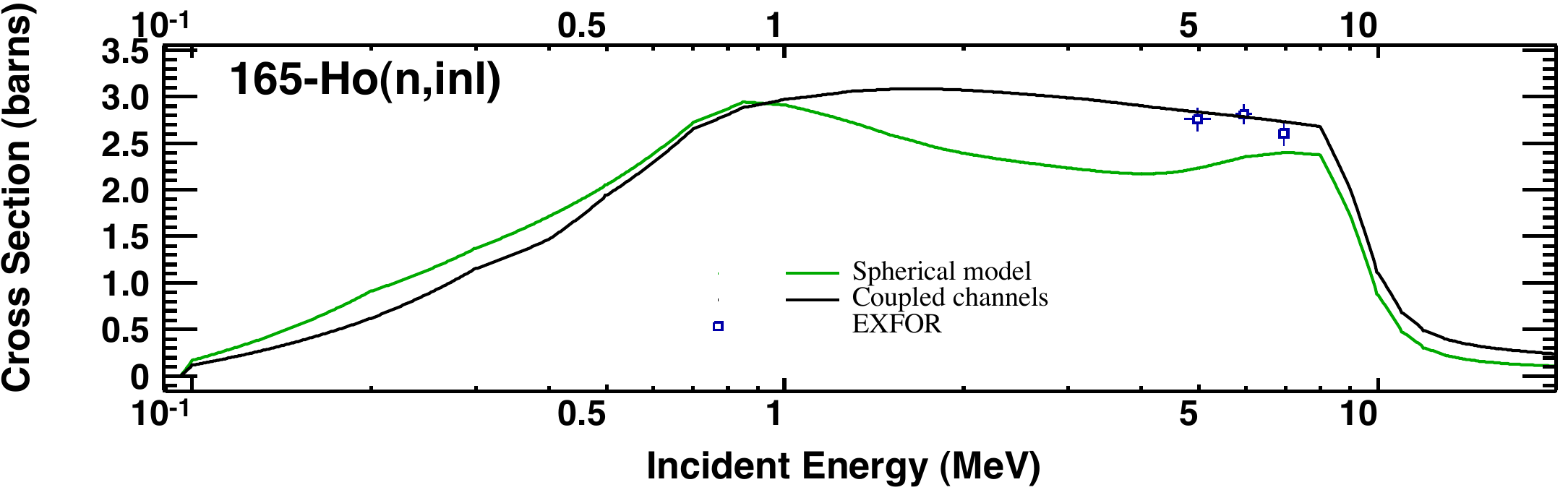}
  \caption{Angle-integrated inelastic cross sections for $^{165}$Ho. Black curves correspond to predictions by our coupled-channel model while green curves were obtained by spherical model calculations. Experimental data taken from EXFOR \cite{EXFOR}.}
  \label{Fig:Inel}
\end{figure}

Fig. \ref{Fig:Inel} shows the total inelastic cross section in the case of neutrons scattered by a $^{165}$Ho target. Again, our coupled-channel model describes well the observed experimental data.

\begin{figure}
\centering
\includegraphics[height=.3\textheight,clip, trim= 5mm 3mm 5mm 14mm]{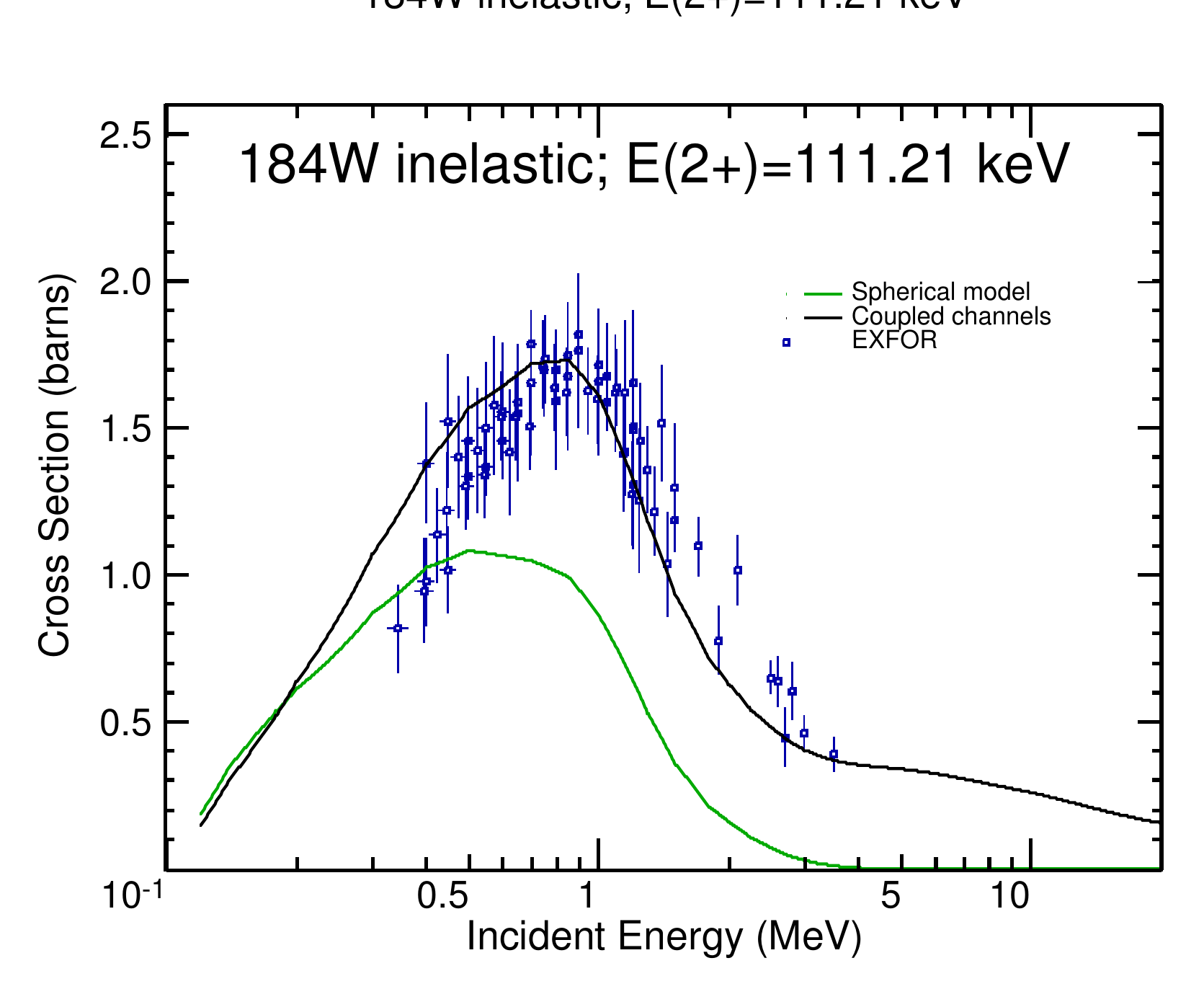}
\caption{Angle-integrated inelastic cross sections for the 2$^+$ state (excitation energy of 111.2~keV) of  the target $^{184}$W, excited on a neutron-induced reaction. The black curve corresponds to coupled-channel calculations within our model, while the green curve indicates, for comparison purposes, the result from a spherical model calculation, using Hauser-Feshbach model to describe the excitation. Experimental data taken from EXFOR \cite{EXFOR}.}
\label{Fig:Inel2+}       
\end{figure}

We also obtain a very good agreement with experimental data for inelastic cross sections for individual excited states. As an example, we show in Fig.~\ref{Fig:Inel2+} the predictions of our model for the angle-integrated inelastic cross section of the first inelastic channel of the target $^{184}$W, which is a 2$^+$ state at excitation energy of 111.2~keV, as a function of the neutron incident energy. We observe again that we are able to achieve a very good description of measured data within our couple-channel model. 

\section{Angular Distributions}

A more careful analysis of differential cross-section experimental data proved necessary due to the large amount of angular distribution data available in the literature, and also because some measurements do not contain pure elastic isotopic data. It is quite common for experiments measuring elastic angular distributions for  rare-earth nuclei to be unable to separate inelastic contributions due to the low-lying excitation energies of their rotational states. In such cases, measured data correspond actually to ``quasi-elastic'' angular distributions, and the  calculated elastic and inelastic differential cross sections have to added up together accordingly for appropriate comparison. In addition, some experiments were done using the natural form of the element, rather than the isotope-specific one.

For these reasons, application of the coupled-channel model for angular distributions was focused on three elements only: Gadolinium, Holmium, and Tungsten. Those three elements were  chosen because the lighter and heavier ones are close to the border of the rare-earth region, while the other is roughly in the middle.

\subsection{Volume Conservation}
 
When an originally spherical configuration assumes a deformed shape, defined by quadrupole and hexadecupole
deformation parameters $\beta_2$ and $\beta_4$, respectively, the volume and densities are not conserved. In Ref.~\cite{Bang:1980}, a method
to ensure volume conservation was proposed, corresponding to applying  a correction to the reduced radius $R_0$, of the form:
\begin{equation}
 R'_{0}=R_0\left(  1-\sum^{}_{\lambda}{\beta_{\lambda}^{2}/4\pi}\right) ,
\label{Eq:radius}
\end{equation}
where terms of the order of $\beta_{\lambda}^{3}$ and higher have been discarded. Ref.~ \cite{NobreND2013} tested the effects of such correction, showing that it is not negligible and seems to bring the integral and differential cross-section calculations to a slightly better agreement with
the experimental data. Therefore, in the following calculations of angular distributions, we decided to implement the radial corrections calculated from Eq.~\ref{Eq:radius}, as it should correspond to a more realistic modeling of the deformed nuclei.

\subsection{Initial results}

In this work we present preliminary results for quasi-elastic differential cross sections for the $^{165}$Ho target. In Ref.~\cite{Nobre-RTFNB} one can also find preliminary results  within the same coupled-channel model of angular distributions for $^{158}$Gd and $^{184}$W.

The ground state of $^{165}$Ho has spin and parity corresponding to 7/2$^-$. As an odd nucleus, its rotational band does not follow the standard 0$^+$, 2$^+$, 4$^+$, etc., scheme. For coupling purposes, it was considered to belong to the ground state rotational band the successive negative-parity states with a difference of spin equal to 1 relative to the ground state, i.\ e., 7/2$^-$, 9/2$^-$, 11/2$^-$, 13/2$^-$, etc. Couple-channel calculations were performed coupling up to the 23/2$^-$ state. The values adopted for the deformation parameters were 
$\beta_{2}=0.3$ 
and $\beta_{4}=-0.020$ \cite{Smith:2001}.

\begin{figure}
\centering
\includegraphics[height=.4\textheight,clip, trim= 9mm 65mm 3mm 62mm]{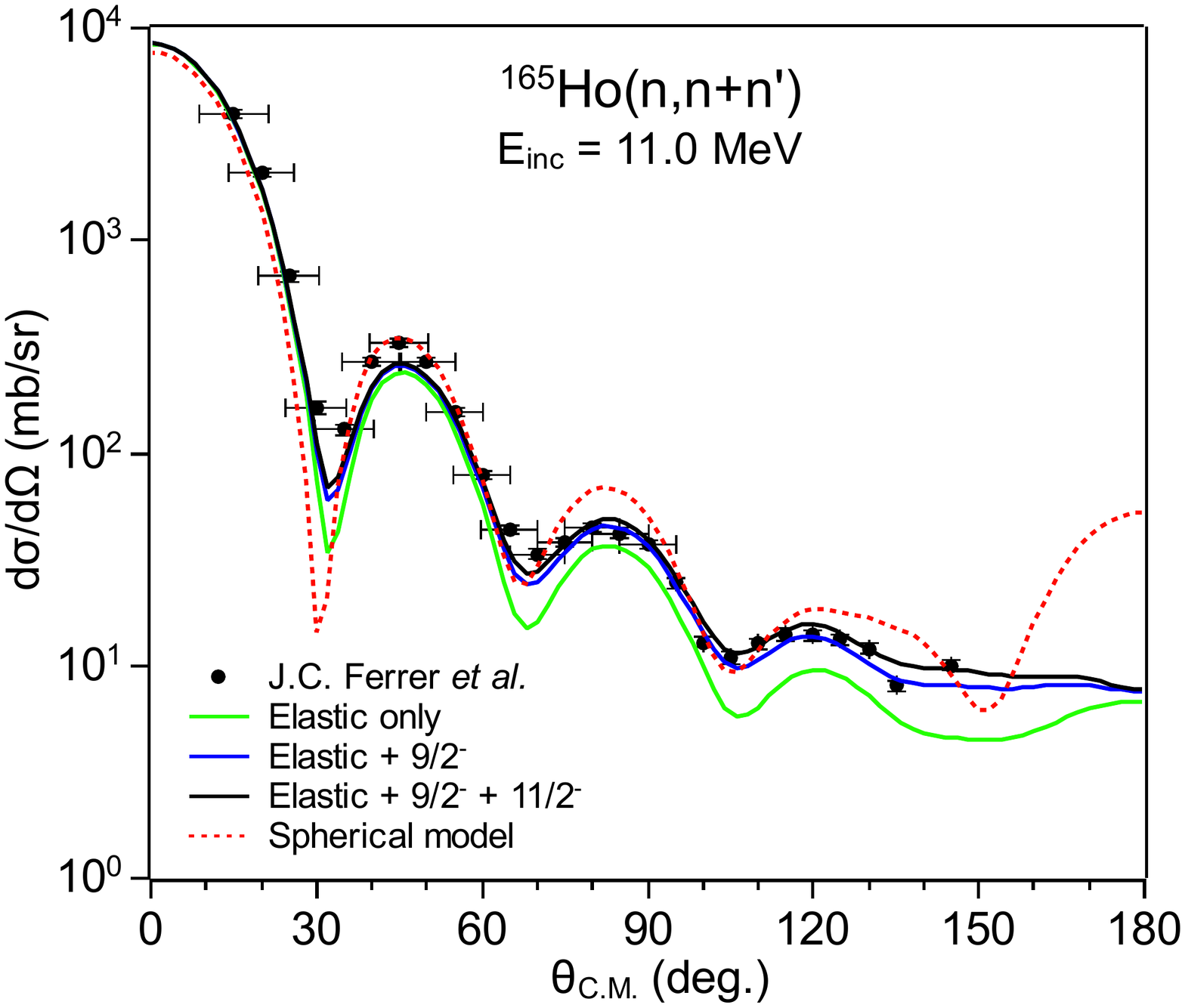}
\caption{Quasi-elastic angular distribution for the neutron-induced reaction on $^{165}$Ho. The green curve corresponds to the results for the elastic channel  only, obtained within our couple-channel model. The blue and black curves contain, in addition to the elastic differential cross section, contributions from the first, and first and second inelastic channels, respectively. For comparison purposes, we also plot the result from a spherical model calculation, as the dashed-red curve. Experimental data taken from  Ref.~\cite{EXFOR}, corresponding to  measurements from Ref.~\cite{Ferrer:1977}.}
\label{Fig:Ho165-Ferrer}       
\end{figure}

As an example,  Fig.~\ref{Fig:Ho165-Ferrer}  presents the predictions of our model when attempting to describe the elastic angular distribution data for $^{165}$Ho, at the neutron-incident energy of 11 MeV, as measured by Ferrer \emph{et al.} \cite{Ferrer:1977}. Actually, a careful analysis of Ref.~\cite{Ferrer:1977} indicates that in that experiment it was not possible to separate the elastic channel from the inelastic ones. Therefore, the data points in Fig.~\ref{Fig:Ho165-Ferrer} should contain inelastic contributions. It is seen Fig.~\ref{Fig:Ho165-Ferrer} that the predictions of our couple-channel model for the elastic angular distribution (green curve) lies consistently below the experimental data. However, when the contribution from the first inelastic state, which is a 9/2$^-$ state (excitation energy of 94.7~keV) is added (blue curve),  the couple-channel prediction approaches the observed cross sections. When the second inelastic state (11/2$^-$ state lying at 209.8 keV) is further added  (black curve), we achieve a very good description of the observed quasi-elastic angular distribution. For comparison purposes we plot on the same figure the result obtained from spherical-model calculations, as the dashed red curve.

\section{Conclusion}

In this work, we demonstrated that we can use the spherical Koning-Delaroche optical potential 
in coupled channels calculations by simply deforming it and making no further modification.  We found that 
we achieved encouraging results in the description of neutron-induced reactions on the rare-earths 
despite the fact that this potential was not designed to describe reactions on deformed nuclei. 
We studied the effect of reducing the radius to ensure volume conservation when deforming an the 
original spherical configuration.  This correction was found to produce small but significant effects 
which improved the agreement with experimental data for the cases we tested.
With our approach, we described experimental data not only for optical-model observables (such as total cross sections,
elastic and inelastic angular distributions), but also for those obtained through compound-nucleus formation (such as
total elastic and inelastic, capture cross sections).  Our results are consistent with the insight that the 
scattering is very close to the adiabatic limit as shown in Ref. \cite{Dietrich:2012}. Although imperfect, 
this simple method is a consistent and general first step towards an optical potential
capable of fully describing the rare-earth region and fills the need of an optical model potential in this important region.

\section*{Acknowledgments}
The work at Brookhaven National Laboratory was sponsored by the Office of Nuclear
Physics, Office of Science of the U.S. Department of
Energy under Contract No. DE-AC02-98CH10886 with
Brookhaven Science Associates, LLC.

%

\bibliography{CC_CNR}
%
%
%
%

\end{document}